\documentclass[aps,prc,twocolumn,groupedaddress]{revtex4}
\usepackage{graphicx,epsfig,amsfonts,amssymb,amsbsy}
\usepackage{graphicx,color}
\usepackage{longtable}
\def\Vec#1{\mbox{\boldmath $#1$}}
\def\vc#1{\mbox{\boldmath $#1$}}
\begin{document}
\title{Concepts of $\alpha$-particle condensation}

\author{Y.~\textsc{Funaki}$^1$, H.~\textsc{Horiuchi}$^{2,3}$, W.~\textsc{von Oertzen}$^{4,5}$, G.~\textsc{R\"opke}$^6$, P.~\textsc{Schuck}$^{7,8,9}$, A.~\textsc{Tohsaki}$^2$ and T.~\textsc{Yamada}$^{10}$}

\affiliation{$^1$ Institute of Physics, University of Tsukuba, Tsukuba 305-8571, Japan}
\affiliation{$^2$ Research Center for Nuclear Physics (RCNP), Osaka University, Osaka 567-0047, Japan}
\affiliation{$^3$International Institute for Advanced Studies, Kizugawa 619-0225, Japan}
\affiliation{$^4$Hahn-Meitner-Institut Berlin, Glienicker Str.100, D-14109 Berlin, Germany}
\affiliation{$^5$Freie Universit\"at Berlin, Fachbereich Physik, Berlin, Germany}
\affiliation{$^6$ Institut f\"ur Physik, Universit\"at Rostock, D-18051 Rostock, Germany}
\affiliation{$^7$ Institut de Physique Nucl\'eaire, 91406 Orsay Cedex, France}
\affiliation{$^8$ Universit\'e Paris-Sud, F-91406 Orsay-C\'edex, France}
\affiliation{$^9$Laboratoire de Physique et Mod\'elisation des Milieux Condens\'es, CNRS et Universit\'e Joseph Fourier, 25 Av.~des Martyrs, BP 166, F-38042 Grenoble Cedex 9, France}
\affiliation{$^{10}$ Laboratory of Physics, Kanto Gakuin University, Yokohama 236-8501, Japan}

\begin{abstract}
Certain aspects of the recently proposed antisymmetrised $\alpha$ particle product state wave function, or THSR $\alpha$ cluster wave function, for the description of the ground state in $^8$Be, the Hoyle state in $^{12}$C, and analogous states in heavier nuclei, are elaborated in detail. For instance, the influence of antisymmetrisation in the Hoyle state on the bosonic character of the $\alpha$ particles is studied carefully. It is shown to be weak. Bosonic aspects in Hoyle and similar states in other self-conjugate nuclei are, therefore, predominant. Other issues are the de Broglie wave length of $\alpha$ particles in the Hoyle state which is shown to be much larger than the inter-alpha distance. It is pointed out that the bosonic features of low density $\alpha$ gas states have measurable consequences, one of which, that is enhanced multi-alpha decay properties, likely already have been detected. Consistent with experiment, the width of the proposed analogue to the Hoyle state in $^{16}$O at the excitation energy of $E_x=15.1$ MeV is estimated to be very small (34 keV), lending credit to the existence of heavier Hoyle-like states. The intrinsic single boson density matrix of a self-bound Bose system can, under physically desirable boundary conditions, be defined unambiguously. One eigenvalue then separates out, being close to the number of $\alpha$'s in the system. Differences between Brink and THSR $\alpha$ cluster wave functions are worked out. No cluster model of the Brink type can describe the Hoyle state with a single configuration. On the contrary, many superpositions of the Brink type are necessary, implying delocalisation towards an $\alpha$ product state. It is shown that single $\alpha$ particle orbits in condensates of different nuclei are almost the same. It is thus argued that $\alpha$ particle (quartet) antisymmetrised product states of the THSR type are a very promising novel and useful concept in nuclear physics.
\end{abstract}

\maketitle
\section{Introduction}

Recently it has been pointed out that certain states in self conjugate nuclei around the alpha-particle disintegration threshold can be described as product states of $\alpha$ particles, all in the lowest $0S$ state. Considerable theoretical and experimental activity has developed since this idea was first put forward in 2001~\cite{thsr}. In this paper we want to further dwell on the basic foundations and predictions of this concept, since the usefulness of the latter has recently been questioned~\cite{zinner07}.

In Refs.~\cite{roepke98,beyer00,sogo09} it was pointed out that in homogeneous nuclear matter $\alpha$-particle condensation is a possible nuclear phase. Therefore, the above mentioned $\alpha$-particle product states in finite nuclei have been proposed to be related to boson condensation of $\alpha$ particles in infinite matter~\cite{roepke98,beyer00,sogo09}. The infinite matter study used a four particle (quartetting) generalisation of the well known Thouless criterion for the onset of pairing as a function of density and temperature. The particular finding in the four nucleon case was that $\alpha$-particle condensation only can occur at very low densities where the quartets do not overlap appreciably. This is contrary to the pairing case where, in weak coupling situations, the Cooper pairs also may strongly mix. It is interesting to note that the low density condition for quartetting was in the meanwhile confirmed in Ref.~\cite{quartet} with a theoretical study in cold atom physics.

Concepts developed for infinite nuclear matter are of value also to 
interpret properties in finite nuclei and to construct useful 
approximations. As examples, we refer to pairing, two and more body correlations, and one body occupation numbers. Pairing is believed to occur in neutron stars which are 
considered as infinite neutron matter becoming superfluid below a critical 
temperature. Pairing also is a useful concept in many finite nuclei, 
in spite of the fact that nuclei are not macroscopic objects. Therefore, in 
reality, they are only in a fluctuating state and we have to 
project, e.g., a BCS state on a definite number of nucleons. 
In spite of the finiteness of nuclei, the BCS state remains a useful 
approximation for the quantum state. For example, the strong reduction of 
measured moments of inertia of such nuclei compared with the classical 
values are explained as a consequence of superfluidity~\cite{bohr}.

 Alpha-particle condensed states may also be of relevance in finite nuclei.
As already pointed out above, in Ref.~\cite{thsr}, see also \cite{schuck07}, we interpret the Hoyle state, i.e. the $0_2^+$-state at $E_x=7.65$ MeV in $^{12}$C, as a product state of three $\alpha$'s and predict that Hoyle-like states very likely also exist in low density states in heavier $n\alpha$ nuclei, close to the $n\alpha$ disintegration threshold. Examples for $^{16}$O and $^{20}$Ne have been presented in~Refs.~\cite{thsr,tohsaki_nara}, employing the so-called THSR wave function, proposed by Tohsaki, Horiuchi, Schuck, and R\"opke. It is analogous to the (number projected) BCS wave function, replacing, however, Cooper pairs by $\alpha$-particles (quartets). In addition, we showed that pure product states of $\alpha$ particles in the threshold states are realised to about 70 percent~\cite{yamada05,matsumura04,funaki08}. We, thus, define a state of condensed $n\alpha$'s, if in a nuclear state the latter forms in good approximation a bosonic product state, all bosons occupying the lowest quantum state of the corresponding bosonic mean field potential.

 In the present paper we will further investigate the concepts and consequences of the THSR wave function. We will carefully study the effect of antisymmetrisation on the bosonic character of the $\alpha$ particles in the Hoyle state. It will be found that, compared to the ground state, its influence is very weak, however, not negligible and in a sense necessary for the $\alpha$ particle gas state of low density to be formed and stabilised. We also focus on measurable properties that are adequately described in this approximation. Considering special results we show the usefulness of this novel type of $\alpha$ cluster wave function.

 The paper is organised as follows. In Sec.~II, we elaborate on the difference between Brink and THSR type wave functions and demonstrate this first for the most simple case of $^8$Be. Sect.~III is dedicated to a study of the antisymmetrisation effects between the $\alpha$'s in the Hoyle state and in Sec.~IV the de Broglie wave length of $\alpha$'s in the Hoyle state is studied. In Sec.~V we discuss some measurable consequences of the bosonic character of $\alpha$'s in compound states, Sec.~VI considers decay properties of Hoyle-like states, and Sec.~VII reveals the similarity of $\alpha$ wave functions in Hoyle-like states in nuclei with different numbers of $\alpha$ particles. Finally in Sec.~VIII, we conclude.

\section{Delocalised alpha-particle condensate vs localised Brink type wave functions}

In Ref.~\cite{zinner07}, it is claimed that the proposed THSR wave function for the description of loosely bound $\alpha$ particle states is an approximation to existing nuclear $\alpha$-cluster states. The authors essentially have in mind localised cluster states of the Brink type~\cite{brink}. In this section we will show in detail, presenting new investigations, that the situation is just the contrary: states which have well born out $n\alpha$-cluster structures like, e.g. the ground and low lying states of $^8$Be and the Hoyle state in $^{12}$C are much more adequately described by THSR than by Brink wave functions.

Wave functions for self conjugate light nuclei which incorporate $\alpha$-cluster substructures have by now been in use in nuclear physics since about half a century~\cite{spl52}. Two $n\alpha$ nuclei have been on the forefront of the investigations: $^8$Be and $^{12}$C, with two and three alpha particles, respectively. The starting point always has been practically the same, that is for the individual alpha particles an intrinsic translationally invariant mean field wave function of Gaussian type, representing the free space alpha-particle wave function is taken, whereas the center of mass (c.o.m.) motion is either determined from a full variational principle or limited parametrised ans\"atze reflecting certain underlying ideas of the alpha-particle motion have been assumed. Let us, therefore, write for an $n\alpha$ nucleus the typical following alpha-cluster wave function
\begin{eqnarray}
&&\hspace{-1cm}\Phi_{n\alpha}({\Vec r}_{1,1}, \cdots, {\Vec r}_{n,4}) = {\cal A}[ \chi({\Vec R}_1, {\Vec R}_2,\cdots, {\Vec R}_n)\nonumber \\
&& \times \phi_{\alpha_1}(\Vec{r}_{1,1},\cdots,\Vec{r}_{1,4})\cdots \phi_{\alpha_n}(\Vec{r}_{n,1},\cdots,\Vec{r}_{n,4})], \label{eq1}
\end{eqnarray}
with $\cal A$ the antisymmetriser, ${\Vec R}_i$ the c.o.m. coordinates of the $\alpha$-particle, and 
\begin{equation}
\phi_{\alpha_k}(\Vec{r}_{k,1},\cdots,\Vec{r}_{k,4}) \propto \exp\Big[ -\sum_{1\leq i<j \leq4}({\Vec r}_{k,i} - {\Vec r}_{k,j})^2/(8b^2)\Big], \label{eq2}
\end{equation}
the normalised intrinsic wave function of the $\alpha$-particle with $(0S)^4$ shell-model configuration. With $b$ a variational parameter, it is well known that Eq.~(\ref{eq2}) can describe the free $\alpha$-particle quite well. 
The $\alpha$-particle wave functions in Eq.~(\ref{eq1}) are fixed to their free space form. The wave function $\chi$ for the c.o.m. motion of the $\alpha$'s with ${\Vec R}_i = \frac{1}{4}({\Vec r}_{i,1}+{\Vec r}_{i,2}+{\Vec r}_{i,3}+{\Vec r}_{i,4})$ is, of course, also chosen translationally invariant, that is it depends only on the relative coordinates ${\Vec R}_{ij}={\Vec R}_i-{\Vec R}_j$ or on the corresponding Jacobi coordinates. The spin-isospin part in Eq.~(\ref{eq1}) is not written out but supposed to be of scalar-isoscalar form. We will not mention it hitherto. As already pointed out, the c.o.m. wave function is either determined from a full variational calculation, minimising the energy, or one adopts a restricted variational ansatz. A famous example is the so-called Brink wave function which places the $\alpha$-particles at certain positions in space~\cite{brink}. For example in the case of $^8$Be, this is
\begin{equation}
\chi_{\scriptsize \Vec{R}}^{\rm Brink}({\Vec R}_1,{\Vec R}_2) \propto \exp\Big[-\Big({\Vec R}_1-{\Vec R}/2 - ({\Vec R}_2 + {\Vec R}/2)\Big)^2/b^2\Big], \label{eq3}
\end{equation}
with an obvious generalisation to the case of $n\alpha$-particles. Usually in Eq.~(\ref{eq3}) $b$ has the same value as for the free space $\alpha$-particles and then Eq.~(\ref{eq3}) implies that the two $\alpha$'s are placed a relative distance ${\Vec R}$ apart from one another. Though this kind of geometrical, crystal-like view of the cluster structure works well for many cases, for instance, parity-violating $^{12}$C+$\alpha$, $^{16}$O+$\alpha$, and $^{40}$Ca+$\alpha$ structures in $^{16}$O, $^{20}$Ne and $^{44}$Ti, respectively~\cite{carbon,44Ti_ohkubo,44Ti_horiuchi}, and also when additionally neutrons are involved~\cite{itagaki04}, it is on the contrary known since several decades that this picture fails for the description of the famous Hoyle state, i.e. the $0_2^+$ state in $^{12}$C.

Since such basic results of cluster physics may not be common knowledge, we here want to present a study with Brink-type cluster wave functions and also compare it with another variational ansatz for $\chi$ with the diametrically opposite point of view of completely delocalised $\alpha$-particles, namely the THSR wave function cited above ~\cite{thsr}. There, the following form for $\chi$ is adopted
\begin{equation}
\chi_{n\alpha}^{\rm THSR}(\Vec{R}_1,\Vec{R}_2,\cdots,\Vec{R}_n) 
 =\chi_0({\Vec {\cal R}}_1)\chi_0({\Vec {\cal R}}_2) \cdots\chi_0({\Vec {\cal R}}_n), \label{eq:chi}
\end{equation}
with ${\Vec {\cal R}}_i = {\Vec R}_i - {\Vec X}_G$ where $\Vec{X}_G=(\Vec{R}_1+\cdots+\Vec{R}_n)/n$ is the total c.o.m. coordinate, and
\begin{equation}
\chi_0({\Vec {\cal R}})=\exp[-2({\Vec {\cal R}}^2/B^2)],
\end{equation}
that is a Gaussian with a large width parameter $B$ which is of the nucleus' dimension. The product of $n$ identical $0S$ wave functions reflects the boson condensate character, discussed in the introduction. This feature is realised as long as the action of the antisymmetriser in Eq.~(\ref{eq1}) is sufficiently weak. It also is useful to notice that with (\ref{eq:chi}) and (\ref{eq1}), we can write for Eq.~(\ref{eq1})
\begin{eqnarray}
&&\hspace{-1.cm}\Phi_{n\alpha} \rightarrow \langle \vc{r}_{1,1},\cdots,\vc{r}_{n,4}|{\rm THSR}\rangle  = {\cal A}[\psi_{\alpha_1}\psi_{\alpha_2}\cdots \psi_{\alpha_n}], \label{eq:B}
\end{eqnarray}
with
\begin{equation}
|{\rm THSR}\rangle =|{\rm THSR}(B) \rangle \equiv {\cal A}|B \rangle \label{eq:B1}
\end{equation}
and
\begin{equation}
\langle \vc{r}_{1,1}, \cdots , \vc{r}_{n,4}|B\rangle = \psi_{\alpha_1} \psi_{\alpha_2}\cdots \psi_{\alpha_n}, \label{eq:B2}
\end{equation}
where $\psi_{\alpha_i} = \chi_0({\Vec {\cal R}})\phi_{\alpha_i}$ and the definitions Eqs.~(\ref{eq:B1}) and (\ref{eq:B2}) will be useful below. Equations (\ref{eq:B}), (\ref{eq:B1}) and (\ref{eq:B2}) highlight the analogy of the THSR wave function with the number projected BCS wave function for pairing
\begin{eqnarray}
&&\hspace{-1cm}\langle {\Vec r}_{1,1}, \cdots, {\Vec r}_{n,2}|{\rm BCS}\rangle = {\cal A}[\phi_{\rm pair}({\vc r}_{1,1}, {\vc r}_{1,2})\nonumber \\
&&\hspace{3cm}\cdots \phi_{\rm pair}({\vc r}_{n,1}, {\vc r}_{n,2})],
\end{eqnarray}
where $\phi_{\rm pair}({\vc r}_{i,1}, {\vc r}_{i,2})$ is the Cooper pair wave function.

 This type of condensate wave function has known, in the meantime, considerable success, notably with an accurate description of the Hoyle state, proposing it as the first of a series of excited states in $n\alpha$ nuclei with $\alpha$-particle product character. Usually one employs Jacobi coordinates $\Vec{\xi}_i$ and then the THSR ansatz for $\chi$ is given by:
\begin{equation}
\chi_{n\alpha}^{\rm THSR}=\exp\Big(-2\sum_{i=1}^{n-1}\mu_i \frac{\Vec{\xi}_{i}^2}{b^2+2\beta^2} \Big), \label{eq3_5}
\end{equation}
with $\mu_i=i/(i+1)$. A slight generalisation of (\ref{eq3_5}) is possible, taking into account nuclear deformation, see below.

Before discussing the Hoyle state, we want to study $^8$Be in some detail because even this nucleus which is known to have {\it intrinsically} a two-alpha dumbbell structure can very well be described in the laboratory frame with the delocalised THSR wave function. Let us repeat Eq.~(\ref{eq1}) for this particular case\begin{equation}
\Phi_{2\alpha}={\cal A}[\chi({\Vec R}_{12}) \phi_{\alpha_1} \phi_{\alpha_2}], \label{eq4}
\end{equation}
with $\Vec{R}_{12} = \Vec{R}_1-\Vec{R}_2$. Note that (\ref{eq4}) is a fully antisymmetric and translationally invariant wave function in $8-1=7$ coordinates. Minimising for a given Hamiltonian with $N$-$N$ forces and Coulomb force~\cite{spl52}, the ground state energy with respect to $\chi$, leads straightforwardly to a Schr\"odinger type of equation for $\chi$, corresponding to the Resonating Group Method (RGM)~\cite{wheeler,spl62},
\begin{equation}
\langle \phi_{\alpha_1}\phi_{\alpha_2} | {\widehat H}-E | {\cal A}[\chi(\Vec{r}) \phi_{\alpha_1}\phi_{\alpha_2}]\rangle =0. \label{eq5}
\end{equation}
With the usual definition of RGM, this equation is transformed into a standard Schr\"odinger equation for the wave function $\Psi_{2\alpha}(\Vec{r})$ of the relative motion of the two $\alpha$-particles 
\begin{eqnarray}
&&\hspace{-0.5 cm} \int d\Vec{r}^\prime {\widetilde h}(\Vec{r},\Vec{r}^\prime)\Psi_{2\alpha}(\Vec{r}^\prime) = E \Psi_{2\alpha}(\Vec{r}), \label{eq6}\\ 
&&\hspace{-0.5 cm} \Psi_{2\alpha}(\Vec{r})=\int d\Vec{r}^\prime n^{1/2}(\Vec{r},\Vec{r}^\prime)\chi(\Vec{r}^\prime), \\
&&\hspace{-0.5 cm} {\widetilde h}(\Vec{r},\Vec{r}^\prime)=\int d\Vec{r}_1 d\Vec{r}_1^\prime n^{-1/2}(\Vec{r},\Vec{r}_1) h(\Vec{r}_1,\Vec{r}_1^\prime) n^{-1/2}(\Vec{r}_1^\prime,\Vec{r}^\prime), \nonumber \\ \label{eq7}
\end{eqnarray}
where 
\begin{eqnarray}
&&\hspace{-1cm} n(\Vec{r},\Vec{r}^\prime) =\langle \delta(\Vec{R}_{12}-\Vec{r})\phi_{\alpha}^2 | {\cal A} [\delta(\Vec{R}_{12}-\Vec{r}^\prime) \phi_{\alpha}^2] \rangle , \nonumber \\
&&\hspace{-1cm} h(\Vec{r},\Vec{r}^\prime) =\langle \delta(\Vec{R}_{12}-\Vec{r})\phi_{\alpha}^2 |{\widehat H} |{\cal A} [\delta(\Vec{R}_{12}-\Vec{r}^\prime) \phi_{\alpha}^2] \rangle. \label{eq8}
\end{eqnarray}

In Eq.~(\ref{eq8}), ${\widehat H}$ is the microscopic Hamiltonian under consideration and $\phi_{\alpha_1} \phi_{\alpha_2}$ is abbreviated by $\phi_{\alpha}^2$.
\begin{figure}[htbp]
\begin{center}
\includegraphics[scale=0.7]{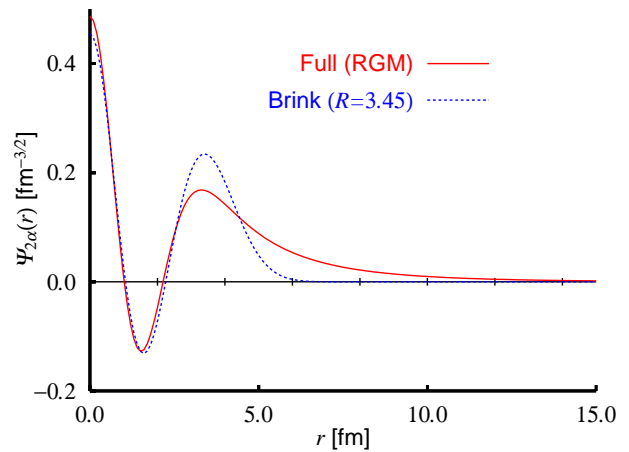}
\caption{Radial parts of wave functions $\Psi_{2\alpha}(\Vec{r})$ for the full RGM solution (full line) and single ``Brink'' component with $R=3.45$ fm (dotted line). The Volkov force No.1 is taken with Majorana parameter value $M=0.56$.}\label{fig:rel_wf}
\end{center}
\end{figure}

\begin{figure}[htbp]
\begin{center}
\includegraphics[scale=0.7]{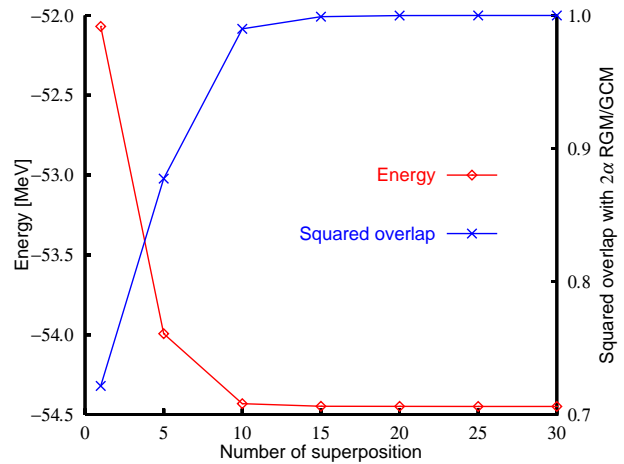}
\caption{Binding energy given by the wave function $\Phi_{2\alpha}$ and the squared overlap between $\Psi_{2\alpha}$ and $\Phi_{2\alpha}$. For $n=1$, a single Brink wave function with optimised $R=3.45$ fm is adopted. See text for definition of $\Phi_{2\alpha}$ and $\Psi_{2\alpha}$. The Volkov force No.1 is taken with Majorana parameter value $M=0.56$.}\label{fig:conv}
\end{center}
\end{figure}

\begin{figure}[htbp]
\begin{center}
\includegraphics[scale=0.7]{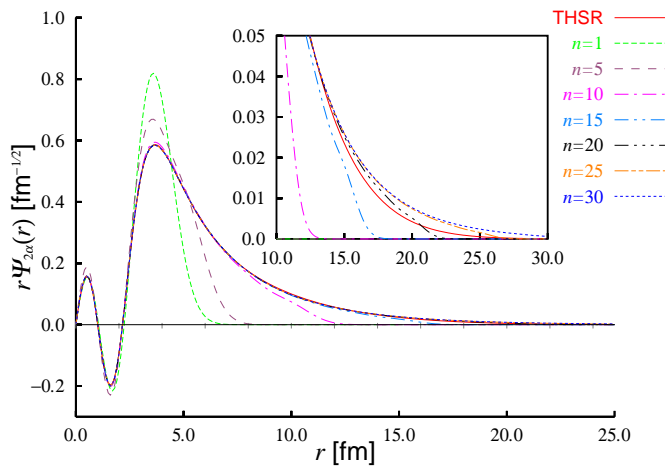}
\caption{Comparison of THSR wave function with single component ``Brink'' wave function $(n=1)$. The convergence rate with the superposition of several $(n)$ ``Brink'' wave functions is also shown (see text for more details).} \label{fig:conv_rel_wf}
\end{center}
\end{figure}

As mentioned, Eq.~(\ref{eq5}) and (\ref{eq6}) have been solved with very high accuracy since 50 years with excellent results for all low energy properties of $^8$Be~\cite{spl52}. The radial part of the wave function $\Psi_{2\alpha}(\Vec{r})$ is shown in Fig.~\ref{fig:rel_wf} by the full line. We see that there exist two nodes, an effect which stems from the Pauli principle. We now will discuss two approximate forms for $\chi(r)$ which are based on the two diametrically opposite views of the nature of $^8$Be already mentioned above: the THSR wave function and the Brink cluster wave function. Let us start with the latter. We have from Eq.~(\ref{eq3})
\begin{equation}
\chi_{R}^{\rm Brink}(r)={\widehat P}^{J=0} \exp \Big[ -\frac{1}{b^2}(\Vec{r}-\Vec{R})^2 \Big], \label{eq9}
\end{equation}
with ${\widehat P}^{J=0}$ the projection operator on $J=0$. In Eq.~(\ref{eq9}) $\Vec{R}$ is a parameter which allows to place the two $\alpha$'s a distance $\Vec{R}$ apart, and $\Vec{r}$ is the relative coordinate between the two $\alpha$'s, i.e. $\Vec{r}=\Vec{R}_1-\Vec{R}_2$. This ansatz seems reasonable, since the microscopic calculation of Ref.~\cite{qmc} indeed indicates that the two $\alpha$'s are about $4$ fm apart. Obviously, the parameter ${\Vec R}$ can be varied to find the optimal position of the $\alpha$-particles. The result of such a procedure is shown on Fig.~\ref{fig:rel_wf} with the broken line, taking the optimal value $R=3.45$ fm ($b$ is kept fixed at its free space value, $b=1.36$ fm). Qualitatively such a ``Brink'' wave function follows the full variational solution (full line). However, in the outer part, for instance in the exponentially decaying tail quite strong differences appear. The squared overlap with the exact solution is 0.722. Of course, such Brink wave functions also can serve as a basis and it is interesting to study the convergence properties. We, therefore, write for the $^8$Be wave function appearing in Eq.~(\ref{eq5})
\begin{equation}
\Phi_{2\alpha}={\cal A}[\chi(r) \phi_{\alpha_1}\phi_{\alpha_2}]=\sum_i f_i \Phi_{2\alpha}^{\rm B} (r,R^{(i)},b) \label{eq10}
\end{equation}
and
\begin{equation}
\Phi_{2\alpha}^{\rm B} (r,R^{(i)},b)={\cal A}[\chi_{R^{(i)}}^{\rm Brink}(r) \phi_{\alpha_1}\phi_{\alpha_2}],
\end{equation}
where the $R^{(i)}$ indicate the various positions of the $\alpha$-particles and $f_i$ are the expansion coefficients. 
The convergence of the squared overlap with the exact solution is studied where we take for the positions $R^{(1)}=1$ fm, $R^{(2)}=2$ fm, $\cdots$, $R^{(n)}=n$ fm. We start with $n=5$. In Fig.~\ref{fig:conv} the convergence rate is shown as a function of $n$ of the squared overlap with the exact solution and same for the energy. The point of $n=1$ is with the optimised single Brink wave function $(R^{(1)}=3.45\ {\rm fm})$. We see that the convergence is not extremely fast but for $n=20$ the squared overlap with the full RGM solution amounts to $0.9999$. Also energy is converged to within $10^{-4}$. In Fig.~\ref{fig:conv_rel_wf} we show the convergence of the wave function $r\Psi_{2\alpha}(r)$. In the insert we see that there is still a slight change in the far tail going from $n=25$ to $n=30$. 

 Let us now investigate the THSR ansatz for $\chi(r)$. There it is assumed from the beginning that the $\alpha$'s are delocalised and a single Gaussian $e^{-r^2/B^2}$ centered at the origin with, however, a large width $B^2=b^2+2\beta^2$, with $\beta$ a variational parameter, is taken. Very much improved results over the single component Brink wave function are obtained. With $\beta=3.24$ fm the squared overlap becomes 97.24\%. However, practically $100$ \% accuracy, compared with the exact solution, can be achieved starting with a slightly improved ansatz, i.e. with an axially symmetric deformed Gaussian which is then projected on the ground-state spin $J=0$ (projections on $J=2,4$ yield the rotational band of $^8$Be)~\cite{funaki_8be},
\begin{eqnarray}
\hspace{-7mm}\chi^{\rm THSR}(r) & = & {\widehat P}^{J=0} \exp \Big( -\frac{r_\perp^2}{b^2+2\beta_\perp^2}-\frac{r_z^2}{b^2+2\beta_z^2} \Big) \label{eq:thsr_8be}\nonumber \\
& \propto &  \frac{\exp( -r^2/B_\perp^2 )}{ir}
 {\rm Erf} \Big(i \frac{(B_z^2-B_{\perp}^2)^{1/2}}{B_{\perp}B_z} r \Big), \label{eq11}
\end{eqnarray}
with $B_i^2=b^2+2\beta_i^2$ and $r_\perp^2=r_x^2+r_y^2$, and ${\rm Erf}(x)$ the error function. The second line of Eq.~(\ref{eq11}) is obtained from a simple calculation. 

Such an intrinsically deformed ansatz is, of course, physically motivated by the observation of the rotational spectrum of $^8$Be indicating a large value of the corresponding moment of inertia. The minimisation of the energy yields $\beta_\perp=\beta_x=\beta_y=1.78$ fm and $\beta_z=7.85$ fm. With these numbers, the squared overlap between the exact $\Psi_{2\alpha}$ and $\Psi_{2\alpha}^{\rm THSR}$ is with $0.9999$ extremely precise~\cite{8be_comments}. In Fig.~\ref{fig:conv_rel_wf} we also show that the THSR wave function agrees almost completely even far out in the tail with the ``exact'' solution with $30$ ``Brink'' components. On the scale of the figure exact and THSR wave functions cannot be distinguished. Let us also mention that beyond $r\sim 3.5$-$4.0$ fm $\chi^{\rm THSR}$ and $\Psi_{2\alpha}$ become practically identical in shape, except for a difference of normalisation, meaning that approximately from the maximum on, the $\alpha$-particles are not influenced any longer by antisymmetrisation and behave as pure bosons.

\begin{figure}[htbp]
\begin{center}
\includegraphics[scale=0.7]{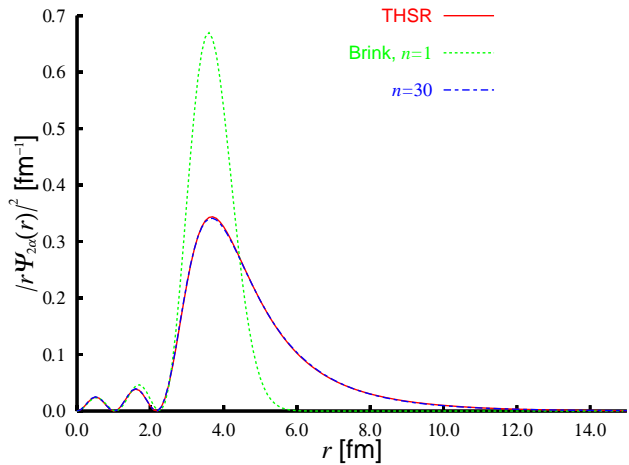}
\caption{Probability distributions $|r\Psi_{2\alpha}(r)|^2$ for the THSR and single component ``Brink'' (denoted as $n=1$) wave functions and for the full RGM solution. On the scale of the figure the two curves of THSR and RGM are indistinguishable. They are normalised as $\int dr |r\Psi_{2\alpha}(r)|^2=1$.}\label{fig:rel_wf_prob}
\end{center}
\end{figure}

As we saw above, the single component, two parameter THSR ansatz, Eq.~(\ref{eq11}), for the relative wave function of two alpha's seems to grasp the physical situation extremely well. The most important part of this wave function is the outer one beyond some 3 fm. There, the two alpha's are in an $S$ wave of essentially Gaussian shape. The corresponding harmonic oscillator frequency is estimated to $\hbar \omega \sim 2$ MeV. Therefore, as long as the two alpha's do not overlap strongly, they swing in a very low frequency harmonic oscillator mode in a wide and delocalised fashion, reminiscent of a weakly bound dimer state. Inside the region $r < 2$-$3$ fm where the two alpha's heavily overlap, because of the strong action of the Pauli principle, the relative wave function has two nodes and small amplitude, as shown in Fig.~\ref{fig:conv_rel_wf}. Contrary to the outer part of the wave function determined dynamically, the behavior of the relative wave function in this strongly overlapping region is determined kinematically, solely reflecting the $r$-dependence of the norm kernel in Eq.~(\ref{eq8}). This is clearly seen from the fact that both THSR and Brink wave functions have very nearly the same behavior in this region. It is also instructive to show the probability $|r\Psi_{2\alpha}(r)|^2$ which is presented in Fig.~\ref{fig:rel_wf_prob}. We see that the latter is practically zero for $r< 2$-$3$ fm, reminiscent of the excluded volume picture which is sometimes adopted when the alpha-particles are treated as pure bosons~\cite{funaki08_prc}. Let us repeat: the alpha's in $^8$Be move practically as pure bosons in a relative $0S$ state of very low frequency as long as they do not come into one another's way, that is as long as they do not overlap. One should stress that this picture holds after projection on good total momentum and good spin, that is in the laboratory frame. It is equally true, as already mentioned, that in the intrinsic frame $^8$Be can be described as a strongly deformed two alpha structure, see ansatz (\ref{eq11}), reminiscent of a dumbbell.

\begin{figure}[htbp]
\begin{center}
\includegraphics[scale=0.3]{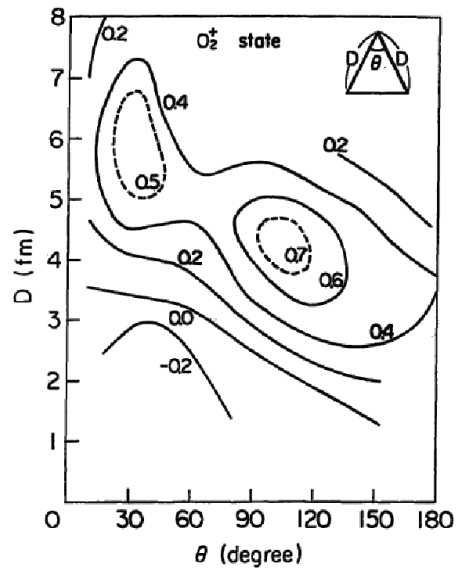}
\caption{Structure of the $0_2^+$ state in $^{12}$C shown by the overlap between the standard Brink cluster wave function of the isosceles configuration and the exact $0_2^+$ wave function. Figure is adopted from Ref.~\cite{uegaki}.}\label{fig:uegaki}
\end{center}
\end{figure}

For the Hoyle state, it is known since long that the situation is qualitatively similar with, however, a much reduced action of the Pauli principle (see discussion below). In the work by Uegaki {\it et al}.~\cite{uegaki,carbon} a contour plot of the overlap between a Brink type wave function and the full RGM solution for $\chi$ of Eq.~(\ref{eq1}) is shown. We reproduce this figure in Fig.~\ref{fig:uegaki}. It is seen that the overlap between the standard cluster wave function and the exact solution is quite poor. In the best case the squared overlap reaches only about 50\%. The authors of that article which dates three decades back, in view of their finding in Fig.~\ref{fig:uegaki}, make the following significant statement to characterise the situation: {\it ``... the $0_2^+$ state has a distinct clustering and has no definite spatial configuration. In other words, $^{12}$C is dissociated in the $0_2^+$ state into weakly interacting three $\alpha$-clusters, which can be considered as a three boson system ...''}. The situation also is highlighted in a recent work by Chernykh {\it et al}.~\cite{chernykh07} where about 55 components of the ``Brink'' type wave functions are needed to reproduce the full RGM solution for the Hoyle state accurately, that is considerably more than in the case of $^8$Be.

In what concerns the THSR wave function for the description of the Hoyle state, the situation is slightly more complicated by the fact that the loosely bound three $\alpha$ configuration is now no longer the ground state but the $0_2^+$ state at $7.65$ MeV excitation energy. We, therefore, have to minimise the energy with the THSR wave function
\begin{equation}
\chi_{3\alpha}^{\rm THSR}=\exp\Big[ -2\sum_{i=1}^2 \mu_i \Big( \frac{\xi_{i\perp}^2}{b^2+2\beta_\perp^2}+ \frac{\xi_{iz}^2}{b^2+2\beta_z^2} \Big) \Big], \label{eq12}
\end{equation} 
where the $\Vec{\xi}_{1,2}$ are the two Jacobi coordinates, and $\mu_1=1/2$, $\mu_2=2/3$, under the condition that the $3\alpha$ wave function $\Phi_{3\alpha}$ is orthogonal to the ground state. 
$\Phi_{3\alpha}$ to be used for the minimisation of the width parameters can schematically be written as 
\begin{equation}
\Phi_{3\alpha}^{\rm THSR} \propto {\widehat P}^{J=0} {\widehat P}_\perp^{\rm (g.s.)} {\cal A}\Big[ \chi_{3\alpha}^{\rm THSR} \phi_\alpha^3 \Big], \label{eq13}
\end{equation}
where ${\widehat P}_\perp^{\rm (g.s.)}$ keeps (\ref{eq13}) orthogonal to the ground state configuration, i.e. ${\widehat P}_\perp^{\rm (g.s.)}=1-|0_1^+\rangle \langle 0_1^+|$. The wave function thus obtained has $99.3$\% squared overlap~\cite{12C_comments} with the full RGM solution of Kamimura {\it et al}.~\cite{kamimura}. The corresponding width parameters have the following values $\beta_\perp=5.3$ fm and $\beta_z=1.5$ fm~\cite{12C_comments}. It should be pointed out that the THSR wave function Eq.~(\ref{eq12}) is again of Gaussian type with a wide extension, centered at the origin. It is completely different from a Brink type wave function with the three $\alpha$-particles placed at definite values in space. A slight improvement of Eq.~(\ref{eq13}) can still be achieved in taking the $\beta_i$ parameters as Hill-Wheeler coordinates and superpose a couple of wave functions of the type (\ref{eq13}) with different width parameters. Practically $100$\% squared overlap with the wave function of the full RGM result of Kamimura {\it et al}.~\cite{kamimura} is then achieved, as documented in \cite{funaki03}. It should be pointed out that the superposition of several Gaussians of the type (\ref{eq12}) does not at all change the physical content of the THSR wave function as a wide extended distribution centered around the origin. As we now have three $\alpha$-particles, all in relative $S$-states, one can begin to talk about coherent features, that is all $\alpha$-particles occupying the same $0S$ orbit. Of course, the Pauli principle is acting, however weakly, and the $0S$ orbit is still occupied to over $70$\%, see Refs. \cite{matsumura04,yamada05,funaki08_prc} and discussion in Sec.~VII. 

We also would like to attract the attention of the reader to the following important point: in spite of the fact that the THSR wave function describes $^8$Be very well, it is, of course, clear that no $\alpha$-particle condensate aspect can be present with only two $\alpha$-particles. This is also born out by the fact that in $^8$Be the $\alpha$-particle wave function still features quite strong influence of the Pauli principle with the two nodes seen on Fig.~\ref{fig:rel_wf}. On the other hand, as seen below on Fig.~\ref{fig:1} in Sect.~\ref{sec:similarity}, in $^{12}$C and $^{16}$O, the $\alpha$-particle wave functions in the condensate states have almost pure $0S$ wave character and, thus, the influence of the Pauli principle is much reduced (see Sec.~III) and the bosonic condensate feature born out. This stems from the fact that e.g. in the Hoyle state the $\alpha$-particles are, on average, by about 70 \% (see Fig.~\ref{fig:uegaki}) farther apart than in $^8$Be and also that the Hoyle state has to be orthogonal to the ground state of $^{12}$C whereas the $\alpha$-structure in $^8$Be represents the ground state itself.

Increasing the number of $\alpha$-particles, the full RGM solution is not possible any longer. However, due to the relative simplicity of the THSR ansatz, analogs to the Hoyle state have been found in $^{16}$O, $^{20}$Ne, always situated close to the $n\alpha$ disintegration threshold~\cite{tohsaki_nara,funaki08}. Due to the high agreement with the full RGM results in the $^8$Be and $^{12}$C cases, one can expect that the THSR wave function also gives accurate results for the heavier systems, grasping well the physical situation of loosely bound $\alpha$-particles moving in identical $0S$ orbits. Naturally the condensate aspect is realized the more, the larger the number of $\alpha$'s.

\begin{figure}[htbp]
\begin{center}
\includegraphics[scale=0.55]{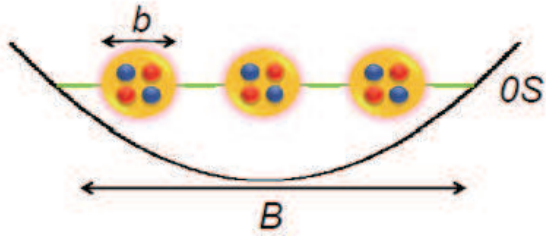}
\caption{Pictorial representation of the THSR wave function for $n=3$ ($^{12}$C). The three $\alpha$-particles are trapped in the $0S$-state of a wide harmonic oscillator $(B)$ and the four nucleons of each $\alpha$ are confined in the $0s$-state of a narrow one $(b)$. All nucleons are antisymmetrised.}\label{fig:cartoon}
\end{center}
\end{figure}

Concluding this section one may just repeat the well known knowledge that the weakly bound $n\alpha$-particle states around the $n\alpha$ disintegration threshold are not at all correctly described by standard $\alpha$-cluster wave functions, with a crystal like structure of the $\alpha$'s. Rather the condensate aspect is dominant and imposes itself as the correct interpretation. For example the Hoyle state, therefore, can be seen as three almost inert $\alpha$-particles moving in their own mean field potential, to good approximation given by a wide harmonic oscillator, whereas the $\alpha$'s are represented by four nucleons captured in narrow harmonic potentials. The situation is given as a cartoon in Fig.~\ref{fig:cartoon}.

\section{Influence of antisymmetrisation}

\begin{figure}[htbp]
\begin{center}
\includegraphics[scale=0.75]{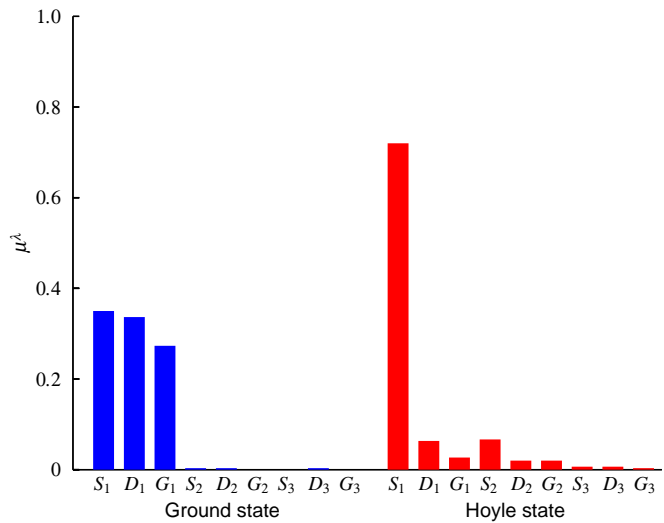}
\caption{Occupation of the single-$\alpha$ orbitals of the Hoyle state of $^{12}$C compared with the ground state~\cite{yamada05}.}\label{fig:occupation}
\end{center}
\end{figure}

As already pointed out in the previous sections, a crucial question is whether for the Hoyle state the THSR wave function (\ref{eq1}) together with (\ref{eq:chi}) can be considered to good approximation as a product state of $\alpha$ particles condensed with their c.o.m. motion into the $0S$ orbital. For this, one has to quantify the influence of the antisymmetriser ${\cal A}$ in (\ref{eq1}). To a certain extent this question was already answered in earlier works~\cite{matsumura04}. For instance in Ref.~\cite{matsumura04} the single $\alpha$ particle density matrix, $\rho(\vc{\cal R},\vc{\cal R}^\prime)$ was constructed and diagonalised for the Hoyle state. The result was that the $\alpha$'s occupy the $0S$ orbit to over 70 percent. The same result was later obtained with the so-called OCM (Orthogonality Condition Model) which is a very well tested method to describe cluster states~\cite{saitoh}. Because of the importance of this result in the present context, we present it again in Fig.~\ref{fig:occupation}.

We see that for the ground state the occupation number distribution agrees with the $SU(3)$ shell model picture (see Ref.~\cite{yamada05}), whereas for the Hoyle state the occupation of the $0S$ c.o.m. wave function of the $\alpha$ particle is, as mentioned, over 70 percent. It also is important to notice that no other state is occupied with more than about 7 percent meaning that the occupation of all other states is down by at least a factor of ten. This is a typical scenario for Bose condensed states.

A more direct way to measure the influence of antisymmetrisation is to consider the following expectation value of the antisymmetriser ${\cal A}$,
\begin{equation}
N(B) = \frac{\langle B|{\cal A}|B\rangle}{\langle B | B \rangle}, \label{eq:exp}\end{equation}
where $|B\rangle$ is the THSR wave function (\ref{eq:B}) without the antisymmetriser, i.e. $|\psi_{\alpha_1}\psi_{\alpha_2}\psi_{\alpha_3} \rangle $, see Eq.~(\ref{eq:B2}). The normalisation of the antisymmetriser ${\cal A}$ is chosen~\cite{tohsaki_KGU} so that $N(B)$ becomes unity in the limit where the inter-cluster overlap disappears, i.e. for the width parameter $B \rightarrow \infty$.

\begin{figure}[htbp]
\begin{center}
\includegraphics[scale=0.725]{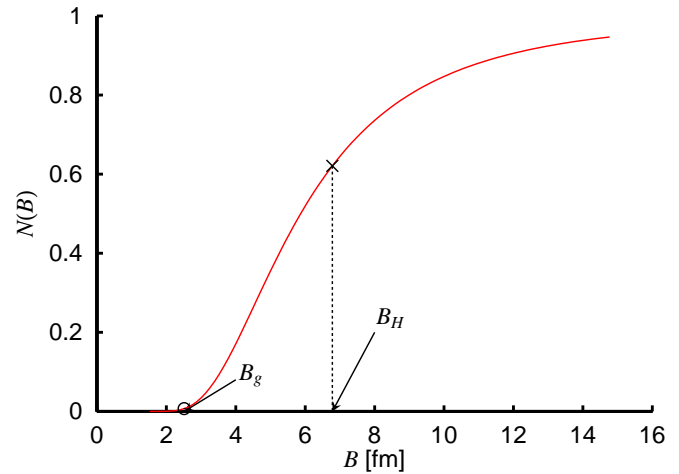}
\caption{The expectation value of the antisymmetriser for the product state $|B\rangle$. The values at the optimal $B$ values, $B_g$ for the ground and $B_H$ for the Hoyle states, are denoted by circle and cross, respectively. See the text for the definition of $|B\rangle$, $B_g$ and $B_H$.}\label{fig:expA}
\end{center}
\end{figure}

In Fig.~\ref{fig:expA}, we show the expectation value $N(B)$, Eq.~(\ref{eq:exp}), of ${\cal A}$ as a function of the width parameter $B$. 
We chose, as optimal values of $B$ for describing the ground and Hoyle states, $B=B_g=2.5$ fm and $B=B_H=6.8$ fm, for which the THSR states best approximate the ground state $|0_1^+\rangle$ and the Hoyle state $|0_2^+\rangle$, respectively, which are obtained by solving the Hill-Wheeler equation. In fact, the normalised THSR state, $|{\rm THSR}(B)\rangle /\sqrt{\langle {\rm THSR}(B)|{\rm THSR}(B) \rangle}$, gives the largest squared overlap 0.93 with the ground state $|0_1^+\rangle$ at $B=B_g$. Similarly, it gives the largest squared overlap 0.78 with the Hoyle state $|0_2^+\rangle$ at $B=B_H$.

We should mention that $B_g\ne b$, since the ground state contains $\alpha$-like correlations which lower the energy with respect to the limit of a pure Slater determinant $(B=b=1.35\ {\rm fm})$ by roughly 5 percent~\cite{thsr,funaki03}. We see from Fig.~\ref{fig:expA} that $N(B_H) \sim~ 0.62$ and $N(B_g) \sim~ 0.007$ what indicates that the influence of antisymmetrisation is strongly reduced in the Hoyle state compared with the one in the ground state.

It is worth to have a closer look to the behavior of $N(B)$. It is seen that first this function raises very steeply, whereas for $B> B_H$ the raise is much slowed down, reflecting the fact that the contribution from the one-nucleon exchange term very slowly fades out~\cite{tohsaki_PTP}.

\begin{figure}[htbp]
\begin{center}
\includegraphics[scale=0.75]{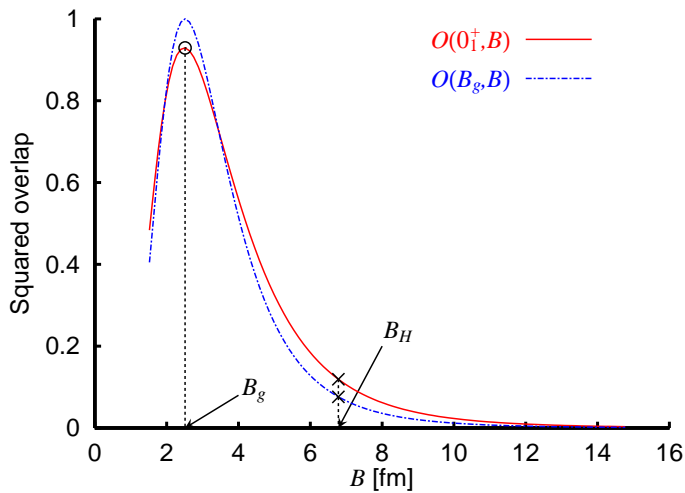}
\caption{The squared overlap of the state $|B\rangle$ with the ground state and $|B_g\rangle$. The values at the optimal $B$ values, $B_g$ and $B_H$ for the ground and Hoyle states, are marked by the circle and cross, respectively. See the text for the definition of $|B\rangle$, $|B_g\rangle$, $B_g$ and $B_H$.}\label{fig:max_ovlp}
\end{center}
\end{figure}

An important point in the present considerations is that the THSR wave function for $B=B_H$ is not automatically orthogonal to the ground state. This is contrary to the situation with condensed cold bosonic atoms for which the density is so low that the overlap of the electron clouds can on average totally be neglected. It is nevertheless interesting to calculate the following overlap of $|{\rm THSR}\rangle$ with the ground state $|0_1^+\rangle$, obtained by solving the Hill-Wheeler equation, or with $|B_g\rangle$, as a function of $B$,
\begin{eqnarray}
&& O(0_1^+,B)=\frac{|\langle 0_1^+ |{\rm THSR} \rangle|^2}{\langle {\rm THSR} | {\rm THSR} \rangle}, \nonumber \\
&& O(B_g,B)\nonumber \\
&&=\frac{|\langle {\rm THSR}(B_g) | {\rm THSR}(B) \rangle|^2}{\langle {\rm THSR} (B_g) |{\rm THSR} (B_g) \rangle \langle {\rm THSR} (B) | {\rm THSR} (B) \rangle}.\nonumber \\
\end{eqnarray}
From Fig.~\ref{fig:max_ovlp}, we find that for both cases, the overlap is less than 0.12 indicating that orthogonality with ground state is nearly realised.

\begin{figure}[htbp]
\begin{center}
\includegraphics[scale=0.725]{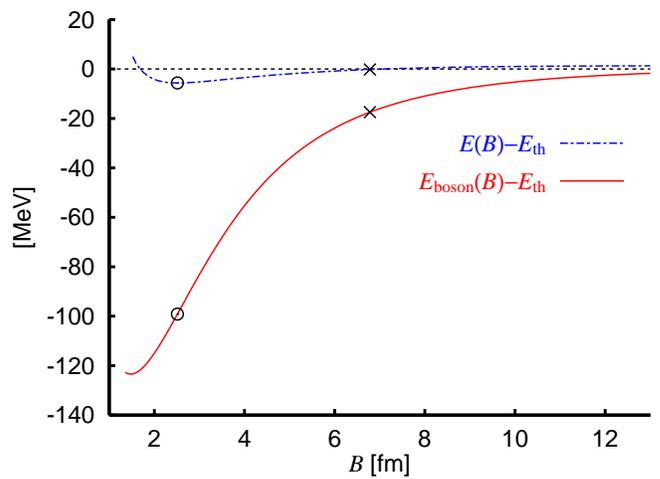}
\caption{The binding energies for $|{\rm THSR}(B)\rangle$ and $|B\rangle$ as a function of $B$, denoted by $E(B)$ and $E_{\rm boson}(B)$, respectively. See the text for the definition of $E(B)$ and $E_{\rm boson}(B)$. They are measured from the calculated $3\alpha$ threshold energy, $E_{\rm th}=3E_{\alpha}=-82.04$ MeV, where $E_{\alpha}$ is the binding energy of the intrinsic $\alpha$ particle. Volkov force No.2~\cite{volkov} with Majorana parameter $M=0.59$ and $b=1.35$ fm are adopted as used in Refs.~\cite{funaki03,kamimura}. The values at the optimal $B$ values, $B_g$ and $B_H$ for the ground and Hoyle states, are marked by the circle and cross, respectively.}\label{fig:eng3a}
\end{center}
\end{figure}

In Fig.~\ref{fig:eng3a}, the energy curves for the THSR wave function with and without the antisymmetriser are shown, where
\begin{eqnarray}
&&E(B)=\frac{\langle {\rm THSR}|H|{\rm THSR} \rangle}{\langle {\rm THSR}|{\rm THSR} \rangle} = \frac{\langle B|H{\cal A}|B \rangle}{\langle B|{\cal A}|B \rangle}, \nonumber \\ \label{eq:E}
&&E_{\rm boson}(B)=\frac{\langle B |H| B \rangle}{\langle B|B \rangle}.
\end{eqnarray}
They are measured from the $3\alpha$ threshold energy, $E_{\rm th}=3E_{\alpha}=-82.04$ MeV, where $E_{\alpha}$ is the binding energy of the intrinsic $\alpha$ particle~\cite{kamimura,funaki03}, obtained with the use of Volkov force No.2~\cite{volkov}. The second equality for $E(B)$ in Eq.~(\ref{eq:E}) holds due to the relation, $[H,{\cal A}]=0$. The minimum for $E(B)$ is given at $B=B_g$, which corresponds to the ground state. The minimum energy $E(B_g)-E_{\rm th}=-5.64$ MeV, as also shown in Ref.~\cite{funaki03}. On the other hand, $E_{\rm boson}(B)-E_{\rm th}\sim -100$ MeV, gives unphysically large binding at small $B$ values around $B_g$, indicating that antisymmetrisation plays an important role for the ground state. As $B$ increases, however, the energy drastically gets smaller, and for $B=B_H$ we have $E_{\rm boson}(B)-E_{\rm th}=-17.5$ MeV. This means that, compared with the ground state at $B\sim B_g$, the effect of antisymmetrisation is much reduced for the Hoyle state. It is, however, still essential to get the energy back on the spot.

\begin{figure}[htbp]
\begin{center}
\includegraphics[scale=0.725]{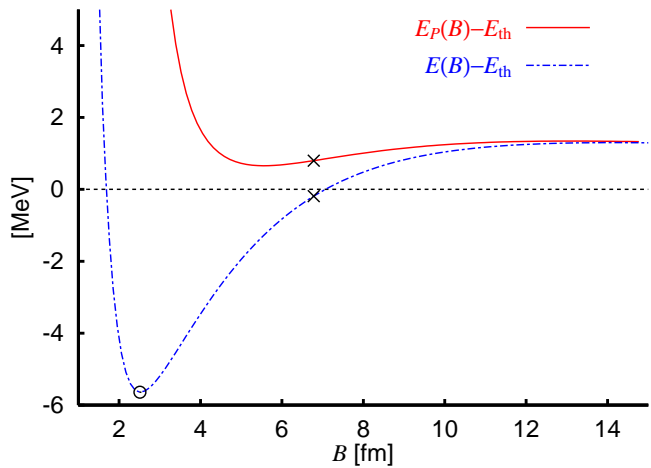}
\caption{
The binding energy in the orthogonal space to the ground state, denoted by $E_P(B)$, together with $E(B)$ in the previous figure. See the text for the definition of $E_P(B)$. They are measured from the calculated $3\alpha$ threshold energy, $E_{\rm th}=-82.04$ MeV. Volkov force No.2~\cite{volkov} with Majorana parameter $M=0.59$ and $b=1.35$ fm are adopted as used in Refs.~\cite{funaki03,kamimura}. The values at the optimal $B$ values, $B_g$ and $B_H$ for the ground and Hoyle states, are marked by the circle and cross, respectively.}\label{fig:eng3a_P}
\end{center}
\end{figure}

It is very important to point out that in this energy curve $E(B)$  the second minimum corresponding to the Hoyle state is not present. This is because the THSR state with $B=B_H$, $|{\rm THSR}(B_H)\rangle$ still includes the ground-state components of about 10 percent what we have seen in Fig.~\ref{fig:max_ovlp}. In fact, if we calculate the following binding energy,
\begin{equation}
E_P(B)=\frac{\langle {\widehat P}_\perp^{({\rm g.s.})}{\rm THSR}|H|{\widehat P}_\perp^{({\rm g.s.})}{\rm THSR} \rangle}{\langle {\widehat P}_\perp^{({\rm g.s.})}{\rm THSR}|{\widehat P}_\perp^{({\rm g.s.})}{\rm THSR} \rangle},
\end{equation}
where the explicit orthogonalisation to the ground state is taken into account for the THSR state, with ${\widehat P}_\perp^{({\rm g.s.})}=1-|0_1^+\rangle \langle 0_1^+|$ like in Eq.~(\ref{eq13}), there appears the minimum corresponding to the Hoyle state at $B\sim B_H$, as shown in Fig.~\ref{fig:eng3a_P}. This is also discussed in Ref.~\cite{uegaki} where Brink-type $3\alpha$ wave functions are used. We should also mention that the Hoyle state is much better approximated by $|{\widehat P}_\perp^{({\rm g.s.})}{\rm THSR}(B)\rangle$ than by $|{\rm THSR}(B=B_H)\rangle$, with proper normalisation factors for both states. The former state gives the largest squared overlap with the Hoyle state, $0.91$ for $B=6.1$ fm, which should be compared with $0.78$ for the latter, a value already mentioned above. Thus the small admixture of the ground-state components for the THSR state is never negligible and the explicit elimination by ${\widehat P}_\perp^{({\rm g.s.})}$ plays an essential role to describe the Hoyle state. It is thus true that the effect of the antisymmetrisation is not negligible even for the Hoyle state in a sense that the projection operator ${\widehat P}_\perp^{({\rm g.s.})}$ includes the compact ground-state components which are strongly subject to the antisymmetriser. Nevertheless, it is worth to emphasise that as a result of the explicit orthogonalisation to the ground state, the Hoyle state cannnot have a compact structure but has a dilute density, for which, in the end, the effect of antisymmetrisation is small.

Therefore, let us point out again that the Hoyle state is to good approximation in a product state of three $\alpha$ particles: about 70 percent of the $\alpha$ particles are in the $0S$ orbit. Only less than 30 percent of the three alpha particles are not in the bosonic product state, a fact due to antisymmetrisation.
According to Ref.~\cite{yamada04}, it is expected that in heavier self conjugate nuclei the alpha particles are still further apart because the stronger Coulomb repulsion lowers the Coulomb barrier. Thus even less influence of antisymmetrisation is expected in heavier Hoyle-like states.

\section{De Broglie Wave Length of $\alpha$-Particles in the Hoyle State}\label{de_broglie}

\begin{figure}[htbp]
\begin{center}
\includegraphics[scale=0.3]{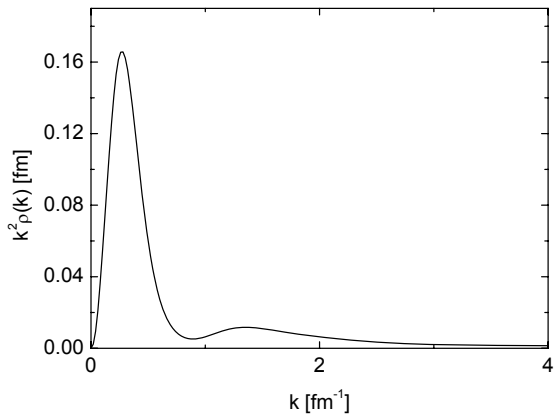}
\caption{Momentum distribution of the $\alpha$ particle in the Hoyle state~\cite{yamada05}.}\label{fig:mom}
\end{center}
\end{figure}

 In this section we show, based on results of detailed microscopic calculations, that the de Broglie wave length $\lambda$ is larger by almost an order of magnitude than the inter $\alpha$ particle distance of about 3-4 fm.

The de Broglie wave length of the $\alpha$'s moving in the Hoyle state can be estimated from the resonance energy of $^8$Be being roughly $100$ keV. Else one can estimate the kinetic energy of the $\alpha$-particles from a bosonic mean field picture using the Gross-Pitaevskii equation~\cite{yamada04}. Figure~3 in Ref.~\cite{yamada04} shows the mean field potential of $\alpha$-particles in the Hoyle state, with an indication of the position of the single $\alpha$ particle energy (180 keV). The kinetic energy of the single $\alpha$ particle is calculated to be 380 keV. From this, the de Broglie wave length $\lambda=h/(2M_\alpha E)^{1/2}$ is, therefore, estimated to be of a lower limit of approximately $20$ fm. A more reliable estimate of the de Broglie wave length is to use the expectation value of $k^2$ for the wave number $k$ of the $\alpha$ particle in the Hoyle state, evaluated from the momentum distribution of the alpha particle, $\rho(k)$, in Fig.~\ref{fig:mom}, obtained by a $3\alpha$ OCM calculation~\cite{yamada05}. The result is $\lambda=2\pi/\sqrt{\langle k^2 \rangle}\sim 20$ fm, consistent with the previous value. These estimates all indicate that the de Broglie wave length is much longer than the inter $\alpha$-particle distance, favoring a mean field approach. 

All these facts make the Bose aspects of the $\alpha$'s in the sense defined in the introduction, plausible and they may reveal specific features of coherence as implied by the notion of a bosonic product state. In the case of nuclear pairing where the number of bosonic constituents, namely the Cooper pairs, is finite and not much larger than the number of $\alpha$'s in light self conjugate nuclei, the collective properties of the $0^+$ ground states and the coherence of these states have been revealed experimentally very early. The most conspicuous example being the strong reduction of the moment of inertia of superfluid deformed nuclei from its classical value and the strong enhancement of the two-neutron transfer to the ground states~\cite{bohr,pair1}. The even-odd staggering in the nuclear masses reveals pairing but not necessarily coherence properties of the pairing state. All these effects of superfluidity are difficult to put into evidence for $\alpha$-particle condensates for the simple reason that  they are resonances around the $\alpha$ disintegration threshold with a finite life time. However, instead of pair-transfer, one can observe $\alpha$-decay. Similar to two-neutron transfer, because of phase coherence, once a first alpha particle leaves the nuclear system, the probability that a second, third, ... will be emitted should be enhanced. This is precisely what we want to report in the following.

\section{Measurable Consequences of loosely bound Alpha-Particle states.}

It is of great importance and interest to discuss eventual measurable consequences for the signature of boson-condensates in nuclei. So far practically all measured quantities of the Hoyle state in $^{12}$C have been reproduced with the THSR wave function with rather good accuracy without any adjustable parameter~\cite{funaki03}. For instance it can be affirmed that the Hoyle state has quite dilute density, about only $1/3\sim 1/4$ of the one of the ground state of $^{12}$C. The density of the Hoyle state is about the same, or slightly less than the one of $^8$Be. To illustrate this dramatic effect it is sufficient to state that the density at the origin is only half of the one in the ground state of $^{12}$C~\cite{kamimura}. This fact alone suggests that in the Hoyle state and other similar states in heavier $n\alpha$ nuclei mentioned above, there is enough space that nucleons cluster into three alphas. The latter mostly interact via the $^8$Be resonance.

We now will give new interpretations of older experiments involving multiple $\alpha$-particle decay out of compound states which reveal coherence effects of an $n\alpha$-particle gas. The idea is that in excited states of heavier $N=Z$ compound nuclei above a certain excitation energy a low density state of $\alpha$-particles can be formed, in a mixed phase of fermions and bosons ~\cite{voe06}. Such states can also be formed on top of an inert core like, e.g.  $^{16}$O or $^{40}$Ca. The decay process is expected, in the light of the present considerations, to show special features as exemplified below. In particular the multiple $\alpha$-decay observed does not correspond to results of the Hauser-Feshbach theory of compound nuclear decay.

\begin{figure*}[htbp]
\begin{center}
\includegraphics[scale=0.63]{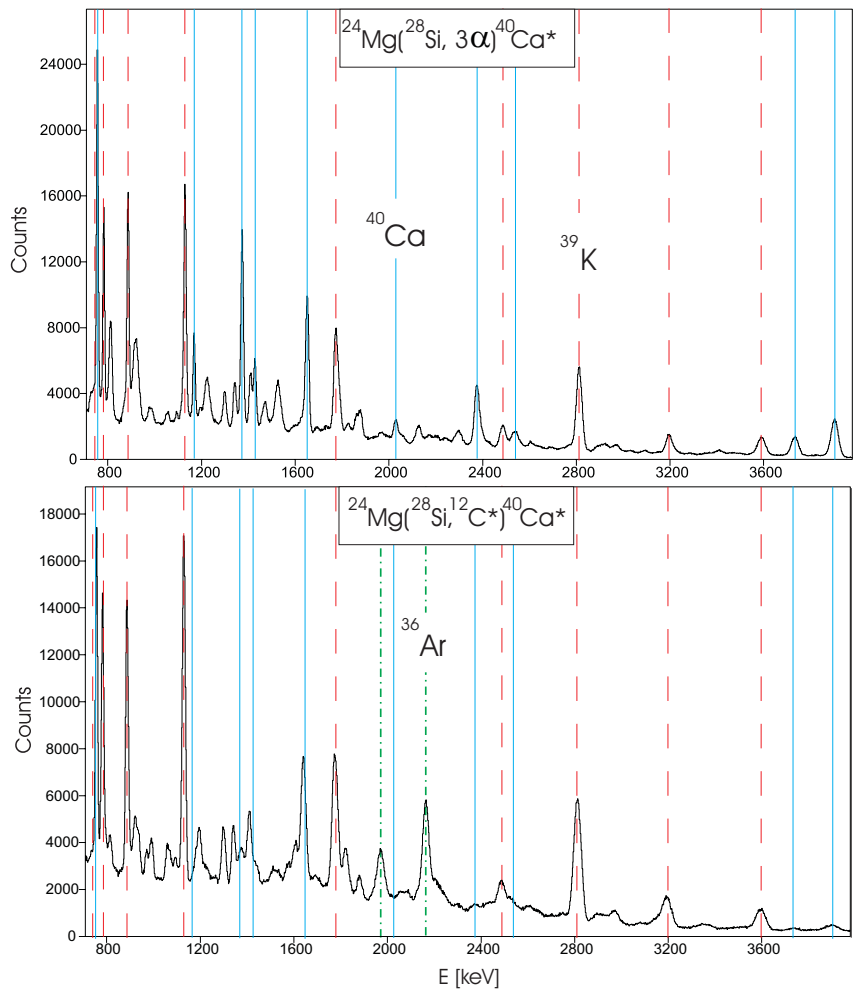}
\caption {Coincident $\gamma$-spectra gated with $\Delta$$E$-$E$-signals with the emission of three random $\alpha$'s in different detectors, in comparison with that obtained by the $^{12}$C$^\ast$($0^+_2$)-gate, as indicated~\cite{Tza05}. The reaction is $^{28}{\rm Si} + $$^{24}{\rm Mg}$ at 130 MeV. Note the additional $^{36}$Ar-line in the lower spectrum, which indicates an additional emission of one $\alpha$. Figure is adopted from Ref.~\cite{Tza05}.} \label{fig:gamc}
\end{center}
\end{figure*}

Previous studies of $^8$Be-emission from excited compound nuclei \cite{vOe00,Tza05,thummerer01} with particle-$\gamma$-coincidences have shown strong effects in the $\gamma$-spectra, if statistical $\alpha$-$\alpha$ emission was compared with the $^8$Be emission.
In these experiments the multiple emission of $\alpha$-particles is registered with the ISIS-$\Delta E$-$E$ particle detection array, in coincidence with the $\gamma$-decay of the residual nuclei registered with the $\gamma$-detector array GASP  at the Laboratorii Nationale di Legnaro (Italy). 
 As an example, we cite the reaction ${^{28}\textrm{Si}} + {^{24}\textrm{Mg}} \rightarrow  {^{52}\textrm{Fe}}$ at 130 MeV~\cite{Tza05} forming a compound nucleus at the excitation energy of $E_x=76$ MeV. For the $\alpha$-$\alpha$-correlations in these experiments an enhancement is observed when they are emitted into the same direction. In view of the  opening angle (27 degr.) of one individual telescope of the ISIS-$\Delta$$E$-$E$-system (42 telescopes), we were able to detect the spontaneous decay of the unbound states just at the decay-thresholds, namely of the $^{8}$Be and the $^{12}$C$^\ast$($0_2^+$) states, into two or three ${\alpha}$-particles, respectively. The decay energies give relative energies of  less than 100 keV and opening angles result, which fit into the solid angle of a single telescope. The corresponding coincident (particle gated) $\gamma$-decays are compared with the spectra obtained from statistical $\alpha$-particle emission into different telescopes, but with the same ${\alpha}$-multiplicity. 

This particularly striking effect is illustrated in Fig.~\ref{fig:gamc} in the case of the above mentioned reaction ${^{28}\textrm{Si}}+{^{24}\textrm{Mg}} \rightarrow {^{52}\textrm{Fe}} \rightarrow {^{40}\textrm{Ca}} + 3{\alpha}$. We compare the $^{12}$C$^\ast$-emission  (lower spectrum shown in Fig.~\ref{fig:gamc}), with that triggered by  the statistical decay with three random  $\alpha$-particles (here the $\gamma$ spectra are dominated by transitions in $^{40}$Ca and $^{39}$K). Quite conspicuous additional gamma rays of  $^{36}$Ar appear in the spectrum gated with $^{12}$C$^\ast$-emission, implying that a 4th ${\alpha}$-particle is emitted, which is not predicted in the Hauser-Feshbach approach for statistical compound decay. 

Further explanation of the observed effect within the concept of a gas of almost ideal bosons in  $^{52}$Fe, (in this case with a  $^{40}$Ca-core), has been proposed in Ref.~\cite{Tza06a}. If such a compound state is formed in the cited reaction, its characteristic feature is a large radial extension with a very large diffuseness of the density distribution. With the large diffuseness of the potential the calculation for the emission of the $^{12}$C$^\ast$ resonance gives a dramatic lowering of the emission barrier, as compared to the statistical multi-$\alpha$-particle emission, it amounts to more than 10 MeV. Thus the residual nucleus ($^{40}$Ca) is populated at much higher excitation energy, and a further $\alpha$-decay can occur. 

The coherent properties of the threshold states consisting of $\alpha$-particles interacting via their resonances~\cite{voe06}, are due to properties of the $^{8}$Be ground state or the  $^{12}$C$^\ast(0^{+}_2)$-state. As reported in Sect.~\ref{de_broglie}, the $\alpha$-particles have a large de Broglie wave length of relative motion, of $\lambda$ = $h/(2M_\alpha E)^{1/2} \sim 20$ fm, or more. In the decay of the compound nucleus the decay steps are usually statistically independent and, thus, the $\alpha$ particles leave the nucleus one by one. However, for the coherent state the $\alpha$'s are already existing (large spectroscopic factors) and their wave functions overlap coherently, the decay may be a simultaneous $n\alpha$ decay, keeping the phase relations of their relative motion. In this way also the Coulomb barrier gets lowered, a fact which further enhances the decay probability. This leads to the observed unbound resonances of $^8$Be and $^{12}$C$^{\ast}(0^+_2)$. This result can be interpreted as the observation of bosonic coherence in the compound nucleus.

 Such enhanced decays also have been observed in studies of  $^8$Be-emission\cite{thummerer00,thummerer01,vOe00} with the corresponding gated $\gamma$-spectra. When compared with the statistical decays into two alpha-particles additional gamma transitions are observed. At the time of these experiments, the discussion of ${\alpha}$-particle coherence was not considered, attempts to explain the observed effects within the extended Hauser-Feshbach (EHF) formalism failed~\cite{thummerer01}.

To summarize we can state that future dedicated experiments, with $\Delta$$E$-$E$-telescopes in coincidence with an efficient $\gamma$-detection array  as described, may be well suited to establish the existence of THSR type states in excited  $N=Z$ nuclei.

\section{Decay properties}

One may ask the question whether Hoyle-like states in nuclei heavier than $^{12}$C can exist. An  argument can be based on the fact that the alpha-particle condensate states occur near the alpha-particle disintegration threshold which rapidly grows in energy with mass and thus the level density in which such a condensate state is embedded raises enormously. For example the alpha-disintegration threshold in $^{12}$C is at $E_x=7.27$ MeV and in $^{16}$O it is already at $E_x=14.4$ MeV. Under ordinary circumstances this could mean that the alpha-particle THSR  state in $^{16}$O, which we suppose to be the well known $0^+$ state at $E_x=15.1$ MeV~\cite{funaki08}, has a very short life time and in Ref.~\cite{zinner07} a Fermi gas estimate is made in this respect. However, on the one hand it is a fact that the supposed $^{16}$O ``Hoyle''-state at $E_x=15.1$ MeV has experimentally, for such a high excitation energy, a startling small width of 160 keV and on the other hand it is easily understandable that such an exotic configuration as four alpha-particles moving almost independently within the common Coulomb barrier, has great difficulties to decay into states lower in energy which all have very different configurations. How else one could explain such a small width of a state this high up in energy? It is precisely one of the strong indications of Hoyle-like states that they should be unusually long lived! It is furthermore well known that the Hoyle state cannot be explained even with the most advanced shell model calculations. Its energy comes at 2-3 times its experimental value~\cite{nocore}. This is a clear indication that shell model configurations only couple extremely weakly to alpha gas states. One can argue that many of the states in $^{16}$O below $E_x=15.1$ MeV are of shell model type. There are also alpha-$^{12}$C configurations but since $^{12}$C also has shell model configuration, it again is difficult for the four alpha condensate state to decay into. 

\begin{table}[hbp]
\begin{center}
\caption{Partial $\alpha$ widths in the $0_6^+$ state of $^{16}$O decaying into possible channels and the total width. The reduced widths defined in Eq.~(\ref{eq:rwa}) are also shown. $a$ is the channel radius.}\label{tab:1}
\begin{tabular}{ccccc}
\hline\hline
 & $^{12}{\rm C}(0_1^+)+\alpha$ & $^{12}{\rm C}(2_1^+)+\alpha$ & $^{12}{\rm C}(0_2^+)+\alpha$ & \raisebox{-1.8ex}[0pt][0pt]{Total} \\
 & ($a=8.0$ fm) & ($a=7.4$ fm) & ($a=8.0$ fm) &    \\
\hline
$\Gamma_L$ (keV) & 26 & 8 & $2\times 10^{-7}$ & $34$  \\
$\theta^2_L(a)$ & 0.006 & 0.004 & 0.15 &   \\
\hline\hline
\end{tabular}
\end{center}
\end{table}

Let us make a more quantitative estimate of the decay width of the $E_x=15.1$ MeV state. Based on the $R$-matrix theory \cite{lane}, the decay width $\Gamma_L$ can be given by the following fomulae:
\begin{eqnarray}
&&\Gamma_L =2P_L(a) \cdot \gamma^2_L(a), \nonumber \\
&&P_L(a)=\frac{ka}{F_L^2(ka)+G_L^2(ka)}, \nonumber \\
&&\gamma^2_L(a)=\theta^2_L(a)\gamma^2_{\rm W}(a), \nonumber \\ 
&&\gamma^2_{\rm W}(a)=\frac{3\hbar^2}{2\mu a^2}, \label{eq:gamma}
\end{eqnarray}
where $k$, $a$ and $\mu$ are the wave number of the relative motion, the channel radius, and the reduced mass, respectively, and $F_L$, $G_L$, and $P_L(a)$ are the regular and irregular Coulomb wave functions and the corresponding penetration factor, respectively. The reduced width of $\theta^2_L(a)$ is related with the wave function $\Psi(0_6^+)$ of the alpha condensate in $^{16}$O obtained as the sixth $0^+$ state in Ref.~\cite{funaki08}, as follows:
\begin{eqnarray}
&&\hspace{-1cm}\theta^2_L(a)=\frac{a^3}{3}{\cal Y}_L^2(a),\nonumber \\
&&\hspace{-1cm}{\cal Y}_L(a)= \Big\langle \Big[ \frac{\delta(r^\prime-a)}{r^{\prime 2}} Y_{L}(\Vec{\hat r}^\prime)\Phi_{L}(^{12}{\rm C}) \Big]_{0} \Big| \Psi(0_6^+) \Big\rangle. \label{eq:rwa}
\end{eqnarray}
where $\Phi_{L}(^{12}{\rm C})$ is the wave function of $^{12}$C, given by the $3\alpha$ OCM calculation~\cite{yamada05}. In Table \ref{tab:1}, we show the partial $\alpha$ decay widths of the $0_6^+$ state $\Gamma_L$ decaying into the $\alpha+^{12}~\!\!{\rm C}(0_1^+)$, $\alpha+^{12}{\rm C}(2_1^+)$ and $\alpha+^{12}{\rm C}(0_2^+)$ channels, total $\alpha$ decay width which is obtained as a sum of the partial widths, and reduced widths $\theta_L^2(a)$ defined in Eq.~(\ref{eq:rwa}). Experimental values are all taken as given by the decay energies. Thus the excitation energy of the calculated $0_6^+$ state is assumed to be 15.1 MeV, the one of the observed $0_6^+$ state. 

\begin{figure*}[htbp]
\begin{center}
\includegraphics[scale=0.6]{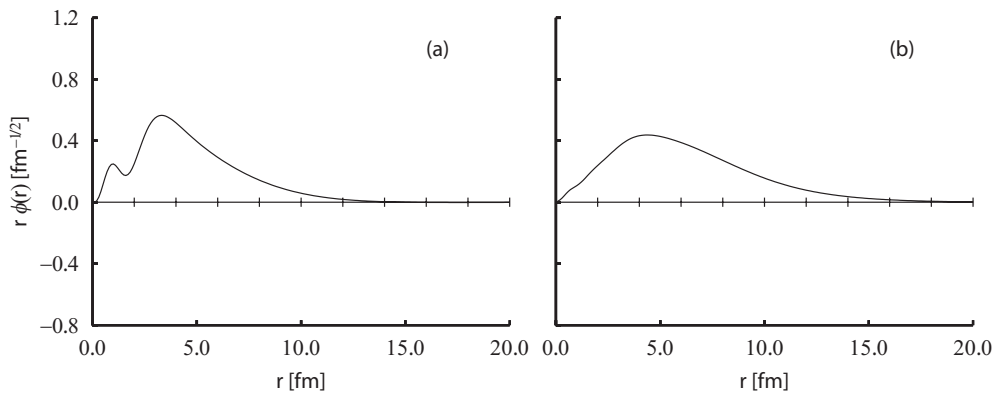}
\caption{Radial distributions of the single-$\alpha$ $S$ orbits, (a) of the $0_2^+$ state in $^{12}$C (Hoyle state) and (b) of the $0_6^+$ state in $^{16}$O.}\label{fig:1}
\end{center}
\end{figure*}

The obtained very small total $\alpha$ decay width of 34 keV, in reasonable agreement with the corresponding experimental value of 160 keV indicates that this state is unusually long lived. The reason of this fact can be explained in terms of the present analysis as follows: Since this state has very exotic structure composed of gas-like four alpha particles, the overlap between this state and $\alpha + ^{12}{\rm C}(0_1^+)$ or $\alpha + ^{12}{\rm C}(2_1^+)$ wave functions with a certain channel radius becomes very small, as this is, indeed, indicated by small $\theta_L^2(a)$ values, 0.006 and 0.004, respectively, and therefore by small $\gamma_L^2(a)$ values. These largely suppress the decay widths expressed by Eq.~(\ref{eq:gamma}) in spite of large values of penetration factors caused by large decay energies 7.9 MeV and 3.5 MeV into these two channels, $\alpha + ^{12}{\rm C}(0_1^+)$ and $\alpha + ^{12}{\rm C}(2_1^+)$, respectively. On the other hand, the decay into $\alpha + ^{12}{\rm C}(0_2^+)$ is also suppressed due to very small penetration caused by very small decay energy 0.28 MeV into this channel, even though the corresponding reduced width takes a relatively large value $\theta_L^2(a)=0.15$, which is natural since the $0_2^+$ state of $^{12}$C has a gas-like three-alpha-particle structure. It is very likely that the above mechanism holds generally for the alpha gas states in heavier $n\alpha$ systems, and therefore such states can also be expected to exist in heavier systems as a relatively long lived resonances.

\section{Similarity of $\alpha$-particle wave functions in Hoyle-like states}\label{sec:similarity}

In Fig.~\ref{fig:1} we show, side by side, radial parts of the single-$\alpha$ $S$ orbits (for definition, see Refs.~\cite{yamada05,funaki08,yamada08}) of the Hoyle state ($^{12}$C) and the $0_6^+$ state in $^{16}$O. We see an almost identical shape. Of course, the extension is slightly different because of the smallness of the system, i.e., we are not dealing with a macroscopic condensate as is discussed above. The nodeless character of the wave function is very pronounced and only some oscillations with small amplitude are present in $^{12}$C, reflecting the weak influence of the Pauli principle between the $\alpha$'s, see the discussion in Sec.~II. On the contrary, due to their much reduced radii, the ``$\alpha$-like'' clusters strongly overlap in the ground states of $^{12}$C and $^{16}$O, producing strong amplitude oscillations which take care of antisymmetrisation between clusters~\cite{yamada05,funaki08}. This example demonstrates the bosonic product nature of the Hoyle state and the $0_6^+$ state in $^{16}{\rm O}$.

\section{Discussions, Summary, and Conclusions}

In this work we considerably deepened several aspects of the THSR description  of low density $\alpha$-particle states in self-conjugate nuclei. We show that the THSR wave function which has alpha-particle condensate structure in analogy with the number projected BCS wave function for Cooper pairs, grasps the physics of loosely bound alpha-particle states as the $^8$Be ground state and the Hoyle state in $^{12}$C, etc. much better than the usual cluster wave functions of the Brink type where rather a crystal structure is involved, the alpha particles with free space extension being placed at certain geometrical positions with respect to one another in the nucleus. Indeed, we have shown for the example of $^8$Be that the superposition of about 30 Brink type wave functions is needed to describe $^8$Be ground state with the same accuracy as the single component THSR wave function which practically coincides with the exact solution of the RGM wave function. Similar results are obtained concerning the Hoyle state, where about $55$ components of the Brink type are needed~\cite{chernykh07}.

One entire section is dedicated to the study of effects from antisymmetrisation between the alpha particles in the THSR (Hoyle) state. We studied the expectation value $N(B)$ of the antisymmetriser as a function of the width parameter $B$ in the THSR wave function which determines the mean distance between the $\alpha$-particles and found that $N(B)$ raises very fast as a function of $B$. For the Hoyle state with $B=B_H$, $N(B_H)$ is about 0.62 whereas for the ground state, i.e. $B=B_g$ we find $N(B_g)$ is about 0.007. So this value increases from the ground state to the Hoyle state by about a factor of hundred, indicating the strongly reduced action of antisymmetrisation in the Hoyle state. Similar conclusions are found for energies and orthonormality relations, see Figs.~\ref{fig:expA}-\ref{fig:eng3a_P}

We also discussed the possible experimental consequences of alpha-particle coherence. This question is more difficult to answer than for nuclear pairing by the fact that those states appear only at low density $(\rho=\rho_0/3\sim \rho_0/4)$ and thus correspond to excited states being of unusual long life time on nuclear scales ($\sim 10^{-17}$~s for e.g. the Hoyle state). Transfer experiments of alpha-particles or measurements of moments of inertia which so clearly demonstrate superfluid features of nuclei in the case of pairing are, therefore, very difficult to conceive in the case of alpha-particles. However, instead of transfer, one may investigate decay. For example in an excited state of $^{52}$Fe, if there exists coherence of a certain number of alpha-particles  on top of an inert core, the simultaneous decay of two or more alpha-particles will be enhanced with respect to a purely statistical decay. Exactly this feature has been observed and related to alpha-particle condensation in the reaction $^{28}$Si $+$ $^{24}$Mg at $E_{\rm lab} = 130$ MeV. However, other experiments may be conceived in the future. For example, with a heavy ion reaction e.g. $^{28}$Si may be (Coulomb) excited to the Ikeda threshold of seven alpha break up, that is $E_x=38.46$ MeV~\cite{ikeda} and then seven alpha-particles may expand as a coherent state verifiable with performant multi-particle detectors~\cite{tanihata}. Measuring energies and angles of the $\alpha$'s may allow to establish an invariant mass spectrum which identifies THSR states, even if they are relatively broad and hidden among other states, seen via gammas or particle evaporation.

Since the alpha-particle condensate states appear around the $n\alpha$ decay threshold, these states are the higher up in the continuum, the heavier the nuclei become. For example in $^{12}$C the Ikeda threshold for alpha decay~\cite{ikeda} is at $E_x=7.27$ MeV, in $^{16}$O it is at $E_x=14.4$ MeV, in $^{24}$Mg at $E_x=28.48$ MeV and so on. One may object that, because of their high excitation energies, those analogs to the Hoyle state will decay very quickly. However, we argued that these alpha gas states have very unusual structure and thus couple only very weakly to states at lower energy. The example of the Hoyle state shows that all states of shell model structure are practically decoupled. This makes up for the large majority of states. For the supposedly analog to the Hoyle state, namely the $0_6^+$ state at $E_x=15.1$ MeV in $^{16}$O, a quantitative estimate of the decay was made, explaining the very small width of $160$ keV.

Going on with this reasoning, it is not at all excluded that alpha gas states in even heavier self conjugate nuclei will have a quite unusual long life time, given their excitation energy high up in the continuum.

Furthermore, the condensate character of the alpha-gas states also has been pointed out in showing, see Fig.~\ref{fig:1}, that the condensate wave function of one alpha-particle changes, apart from a trivial size effect, very little in going e.g. from the Hoyle state in $^{12}$C to the corresponding state at $E_x=15.1$ MeV in $^{16}$O, that is, a condensate wave function is a $0S$ state.

One of our strong arguments that the alpha-particles in the Hoyle state and possibly in the $0_6^+$ of $^{16}$O, form an alpha-particle gas, captured inside the Coulomb barrier, is deduced from the fact that we constructed a single alpha-density matrix whose eigenvalues show that the alpha product states are realised to around $60$-$70$\% , all bosons being in the lowest quantum state~\cite{yamada05,matsumura04,funaki08}. The occupancies of all higher quantum levels are down by at least a factor of ten. However, the authors of~Ref.~\cite{zinner07} mentioned that the way how to define the density matrix in a self bound Bose system with a finite number of particles, is not unique and that different definitions might give different occupancies. We followed in this respect the line of thought outlined by Pethick and Pitaevskii in~Refs.~\cite{pethick00,pita} where they say that if in a homogeneous system there is Bose condensation, then there is no reason that, if the same system is put into an external potential or if the system is self bound in a mean field potential, the system be not also in a non-fragmented condensate state, as long as the intrinsic system is not excited. We showed that our definition of the boson density matrix satisfies this physically very reasonable boundary condition in using Jacobi coordinates for the internal system~\cite{yamada08}. We furthermore showed that for the $0S$ harmonic oscillator wave function the internal one-body density matrix is uniquely determined under another reasonable condition~\cite{yamada09}. The uniqueness for more general wave functions is also demonstrated ~\cite{takahashi,few-body}.

In the light of this finding, we would like to discuss again the content of the THSR alpha-particle condensate wave function~Eq.~(\ref{eq:B}). It is very important to remark, as is explained in Ref.~\cite{thsr}, that  this antisymmetrised $\alpha$ particle product wave function contains two limits exactly. On the one hand, for $B=b$ we have a pure harmonic oscillator wave function because the antisymmetriser generates out of the product of simple Gaussians all higher nodal wave functions of the harmonic oscillator~\cite{mono}. On the other hand, for $B\gg b$ the THSR wave function tends to a pure product state of alpha-particles, i.e. a mean field wave function, since in this case the antisymmetriser can be neglected. Indeed $B$ triggers the extension of the nucleus, i.e.~its average density. For alpha particles kept at their free space size (small $b$), the alpha-particles are then for large $B$-values far apart from one another and do not feel any action from the Pauli principle, see a detailed discussion of the action of the antisymmetriser as a function of density in Ref.~\cite{yamada05}. The question is then whether, e.g. for the Hoyle state, the above wave function is closer to a shell model like Slater determinant or to an $\alpha$-particle product state. Precisely this question is answered by the above discussed eigenvalues of  the density matrix. In this respect it is important to point out that in the calculation of the afore-mentioned density matrix always the total c.o.m. motion has been split off in the wave function of Eq.~(\ref{eq1}) and that for the remaining relative c.o.m. coordinates the Jacobi ones have been used, as is clearly explained in Refs.~\cite{matsumura04,yamada05}. In Refs.~\cite{matsumura04,yamada05} it has been shown, as explained, that the $\alpha$'s in the Hoyle state occupy to over 70~\% the $0S$-orbit. Therefore, the Hoyle state is in good approximation a product of three alpha particles, that is a condensate. This finding also is corroborated by our study on antisymmetrisation effects in the Hoyle state in Sec.~III. As already mentioned, it has been found that the effects of antisymmetrisation are weak.

In summary we can say that our study clearly shows that the loosely bound alpha-particle states of very low density, close to the decay threshold in self-conjugate nuclei, are characterised by a shallow self consistent mean field of wide extension in which the c.o.m. motion of the alpha-particles occupies with about 70\% the lowest $0S$ level. In spite of the very different number of particles and other important differences, the situation has, therefore, some analogy with the case of cold atoms. Avoiding vague and qualitative arguments, we hope to have been sufficiently detailed and convincing, based on new insights from precise numerical results, to confirm the existence of low density bosonic $\alpha$ particle gas states in nuclei and to affirm the usefullness of this novel concept in nuclear physics described with the THSR wave function.

\section*{Acknowledgements}
The authors express special thanks to T. Neff whose detailed studies of the $^8$Be wave function, made available to us, helped, together with extended discussions, our work appreciably. They also thank Kanto Gakuin University (KGU) and Yukawa Institute for Theoretical Physics at Kyoto University, Japan. Discussions during the KGU Yokohama Autumn School of Nuclear Physics and the YIPQS international molecule workshop held in October 2008 were useful to complete this work.


\begin{thebibliography}{99}
\bibitem{thsr}
 A.~Tohsaki, H.~Horiuchi, P.~Schuck and G.~R\"opke, Phys. Rev. Lett. {\bf 87}, 192501 (2001).
\bibitem{zinner07}
N.T. Zinner and A.S. Jensen, Phys. Rev. C {\bf 78}, 041306(R) (2008).
\bibitem{roepke98}
G. R\"opke, A. Schnell, P. Schuck, and P. Nozi\`eres, Phys. Rev. Lett. {\bf 80}, 3177 (1998).
\bibitem{beyer00}
M. Beyer, S.A. Sofianes, C. Kuhrts, G. R\"opke, and P. Schuck, Phys. Lett. {\bf 448B}, 247, (2000).
\bibitem{sogo09}
T.~Sogo, R.~Lazauskas,~G. R\"opke, and P. Schuck, Phys. Rev. C {\bf 79}, 051301(R) (2009).
\bibitem{quartet}
B. Doucot, J. Vidal, Phys. Rev. Lett. {\bf 88}, 227005 (2002); S. Capponi, G. Roux, P. Lecheminant, P. Azaria, E. Boulat, S.R. White, Phys. Rev. A {\bf 77}, 013624 (2008); P. Lecheminant, E. Boulat, and P. Azaria, Phys. Rev. Lett. {\bf 95}, 240402 (2005).
\bibitem{bohr}
A. Bohr and B. R.Mottelson, {\it Nuclear Structure} (Benjamin, New
York, 1975), Vol. 2; P. Ring and P. Schuck, {\it The Nuclear Many Body Problem}
(Springer-Verlag, New York, 1980); J.-P. Blaizot and G. Ripka, {\it Quantum Theory of Finite Systems} (MIT, Cambridge, MA, 1986).
\bibitem{schuck07}
For example, P. Schuck, Y. Funaki, H. Horiuchi, G. R\"opke, A. Thosaki, and T. Yamada, Prog. Part. Nucl. Phys. {\bf 59}, 285 (2007).
\bibitem{tohsaki_nara}
A. Tohsaki, H. Horiuchi, P. Schuck and G. R\"opke, Nucl. Phys. A {\bf 738}, 259 (2004).
\bibitem{yamada05}
T. Yamada and P. Schuck, Eur. Phys. J. A. {\bf 26}, 185 (2005).
\bibitem{matsumura04}
H. Matsumura and Y. Suzuki, Nucl. Phys. A {\bf 739}, 238 (2004).
\bibitem{funaki08}
Y. Funaki, T. Yamada, H. Horiuchi, G. R\"opke, P. Schuck, and A. Tohsaki, Phys. Rev. Lett. {\bf 101}, 082502 (2008).
\bibitem{brink}
D.M. Brink, {\it in Proceedings of the International School of Physics ``Enrico Fermi'', Course {\bf 36}} (Academic Press, New York, London, 1966) p. 247.
\bibitem{spl52}
R. Tamagaki {\it et al}., Prog. Theor. Phys. Suppl. {\bf 52}, 1 (1972), and references therein.
\bibitem{carbon}
For example, Y. Fujiwara, H. Horiuchi, K. Ikeda, M. Kamimura, K. Kat${\rm{\bar{o}}}$, Y. Suzuki, and E. Uegaki, Prog. Theor. Phys. Suppl. {\bf 68}, 29 (1980).
\bibitem{44Ti_ohkubo}
T. Yamaya, K. Katori, M. Fujiwara, S. Kato and S. Ohkubo, Prog. Theor. Phys. Suppl. {\bf 132}, 73 (1998); F. Michel, S. Ohkubo and G. Reidemeister, Prog. Theor. Phys. Suppl. {\bf 132}, 7 (1998).
\bibitem{44Ti_horiuchi}
T. Wada and H. Horiuchi, Phys. Rev. C {\bf 38}, 2063 (1988).
\bibitem{itagaki04}
N. Itagaki, T. Otsuka, K. Ikeda and S. Okabe, Phys. Rev. Lett. {\bf 92}, 142501 (2004).
\bibitem{wheeler}
J. A. Wheeler, Phys. Rev. {\bf 52}, 1083 (1937); {\it ibid.}, 1107 (1937)
\bibitem{spl62}
K. Ikeda, R. Tamagaki, S. Saito, H. Horiuchi, A. Tohsaki-Suzuki and M. Kamimura, Prog. Theor. Phys. Suppl. {\bf 62} (1977).
\bibitem{qmc}
R. B. Wiringa, S. C. Pieper, J. Carlson, and V. R. Pandharipande, Phys. Rev. C {\bf 62}, 014001 (2000).
\bibitem{funaki_8be}
Y. Funaki, H. Horiuchi, A. Tohsaki, P. Schuck and G. R\"opke, Prog. Theor. Phys. {\bf 108}, 297 (2002).
\bibitem{8be_comments}
The $2\alpha$ wave function given in Ref.~\cite{funaki_8be} as the exact $2\alpha$ RGM solution is different from the one given here. We now realized that to a better solution, i.e. to a better energy convergence within a hundred eV, we should also include the components of Brink wave function with values of $R$ as far as 30 fm, while previously the Brink components only up to $R=12.5$ fm have been adopted.
\bibitem{funaki08_prc}
Y. Funaki, H. Horiuchi, G. R\"opke, P. Schuck, A. Tohsaki and T. Yamada, Phys. Rev. C {\bf 77}, 064312 (2008).
\bibitem{uegaki}
E. Uegaki, S. Okabe, Y. Abe, and H. Tanaka, Prog. Theor. Phys. {\bf 57}, 1262 (1977); E. Uegaki, Y. Abe, S. Okabe, and H. Tanaka, Prog. Theor. Phys. {\bf 59}, 1031 (1978); {\bf 62}, 1621 (1979).
\bibitem{chernykh07}
M. Chernykh, H. Feldmeier, T. Neff, P. von Neumann-Cosel, and A. Richter, Phys. Rev. Lett. {\bf 98}, 032501 (2007). 
\bibitem{12C_comments}
The reason why the present value for the squared overlap is different from 0.97 which was given in Ref.~\cite{funaki03} is as follows: In Ref.~\cite{funaki03}, as the ground state configuration used in the projection operator ${\widehat P}_\perp^{({\rm g.s.})}$, we took the wave function which corresponds to the minimum of the energy surface in Fig.~1 of Ref.~\cite{funaki03}. However in the present definition of ${\widehat P}_\perp^{({\rm g.s.})}$, we use the ground-state solution of the Hill-Wheeler equation in the two-parameter space of $\beta_\perp$ and $\beta_z$.
\bibitem{kamimura}
Y. Fukushima and M. Kamimura, {\it{Proc. Int. Conf. on Nuclear Structure}}, Tokyo, 1977, ed. T. Marumori (Suppl. of J. Phys. Soc. Japan, {\bf 44}, 225 (1978)); M. Kamimura, Nucl. Phys. A {\bf 351}, 456 (1981).
\bibitem{funaki03}
Y. Funaki, A. Tohsaki, H. Horiuchi, P. Schuck and G. R\"opke, Phys. Rev. C {\bf 67}, 051306(R) (2003).
\bibitem{saitoh}
S. Saito, Prog. Theor. Phys. {\bf 40}, 893 (1968); {\bf 41}, 705 (1969); Prog. Theor. Phys. Suppl. No. 62, 11 (1977).
\bibitem{tohsaki_KGU}
A. Tohsaki, Y. Funaki, H. Horiuchi, G. R\"opke, P. Schuck and T. Yamada, {\it{in Proceedings of KGU Yokohama Autumn School of Nuclear Physics}}, Yokohama, 2008, ed. T. Yamada and Y. Funaki (Int. J. Mod. Phys. A {\bf 24}, 2003 (2009)).
\bibitem{tohsaki_PTP}
A. Tohsaki-Suzuki, Prog. Theor. Phys. Suppl. {\bf 62}, 191 (1977).
\bibitem{volkov}
A.B. Volkov, Nucl. Phys. {\bf 74}, 33 (1965).
\bibitem{yamada04}
T. Yamada and P. Schuck, Phys. Rev. C {\bf 69}, 024309 (2004).
\bibitem{pair1}
W. von Oertzen and A Vitturi, Rep. Prog. in Phys. {\bf 64}, 1247 (2001).
\bibitem{voe06}
W. von Oertzen, Eur. Phys. J. A {\bf 29}, 133 (2006).
\bibitem{vOe00}
W.~von~Oertzen, Phys. Scr. T \textbf{88}, 83 (2000).
\bibitem{Tza05}
Tz.~Kokalova {\em et al.}, Eur. Phys. J. A {\bf 23}, 19 (2005).
\bibitem{thummerer01}
S.~Thummerer, W. von~Oertzen, B.~Gebauer, S.~Lenzi, A.~Gadea, D.R.~Napoli, C.~Beck and M.~Rousseau, J. Phys. G \textbf{27} 1405 (2001).
\bibitem{Tza06a}
Tz.~Kokalova, N.~Itagaki, W. von Oertzen, and C. Wheldon, Phys. Rev. Lett. {\bf 96}, 192502 (2006).
\bibitem{thummerer00}
S.~Thummerer {\em et al.}, Phys. Scr. T \textbf{88},  114 (2000).
\bibitem{nocore}
B. R. Barrett, B. Mihaila, S. C. Pieper, and R. B. Wiringa, Nucl. Phys. News, {\bf 13}, 17 (2003), and B. R. Barrett, private communication.
\bibitem{lane}
A. M. Lane and R. G. Thomas, Rev. Mod. Phys. {\bf 30}, 257 (1958).
\bibitem{yamada08}
T. Yamada, Y. Funaki, H. Horiuchi, G. R\"opke, P. Schuck, and A. Tohsaki, Phys. Rev. A {\bf 78}, 035603 (2008).
\bibitem{ikeda}
K.~Ikeda, N.~Takigawa, and H.~Horiuchi, Prog. Theor. Phys. Suppl. extra number, 464 (1968).
\bibitem{tanihata}
I.~Tanihata and W. von Oertzen, private communication.
\bibitem{pethick00}
C.J. Pethick and L.P. Pitaevskii, Phys. Rev. A {\bf 62}, 033609 (2000).
\bibitem{pita}
L.P. Pitaevskii, private communication.
\bibitem{yamada09}
T. Yamada, Y. Funaki, H. Horiuchi, G. R\"opke, P. Schuck, and A. Tohsaki, Phys. Rev. C {\bf 79}, 054314 (2009).
\bibitem{takahashi}
Y. Suzuki and M. Takahashi, Phys. Rev. C {\bf 65}, 064318 (2002).
\bibitem{few-body}
Y. Suzuki, W. Horiuchi, M. Orabi, K. Arai, Few-Body Systems {\bf 42}, 33 (2008).\bibitem{mono}
T. Yamada, Y. Funaki, H. Horiuchi, K. Ikeda, and A. Tohsaki, Prog. Theor. Phys. {\bf 120}, 1139 (2008).
\end{thebibliography}
\end{document}